\documentclass[twocolumn, aps, prb, showpacs]{revtex4}
\usepackage{graphicx}
\usepackage{graphics}
\usepackage{amsmath}

\begin{document}

\title{Nonequilibrium critical relaxation in the presence of extended defects}
\author{Andrei A. Fedorenko}
\affiliation{Martin-Luther-Universit\"{a}t Halle-Wittenberg, Fachbereich Physik,
D-06099 Halle, Germany}
\date{December 2, 2004}

\begin{abstract}
We study nonequilibrium critical relaxation properties of systems with
quenched extended defects, correlated in $\varepsilon_d$ dimensions and
randomly distributed in the remaining $d-\varepsilon_d$ dimensions. Using a
field-theoretic renormalization-group approach, we find the scaling behavior of the
nonequilibrium response and correlation functions and calculate the initial slip
exponents $\theta$ and $\theta^{\prime}$, which describe the growth of
correlations during the initial stage of the critical relaxation, in the two-loop
approximation.
\end{abstract}

\maketitle

\section{Introduction}

In recent years, much attention has been attracted by the so-called
short-time critical dynamics. \cite%
{janssen-89,oerding-93,oerding-93-2,kissner-92,oerding-95,chen-01,chen-00,%
chen-02,zheng-98}
A thermodynamic system in a critical point  is characterized by long-range
correlations in equilibrium, so when such a system is quenched from a high
temperature $T_0\gg T_c$ to the critical point $T_c$, the growth of correlations
governs the relaxation process. As shown in Ref.~\onlinecite{janssen-89},
the relaxation process displays a universal scaling behavior already at times
just after a microscopic time scale $t_{\mathrm{mic}}$ needed for the system
to forget its microscopic details and lasts until the system has lost all memory
about its macroscopic initial condition. The initial nonequilibrium stage of the
critical relaxation is characterized by new critical exponents $\theta$ and
$\theta^{\prime}$, which describe the behavior of the response function
$G(t,t^{\prime})\propto(t/t^{\prime})^{\theta}$ for $t^{\prime}\to{0}$ and the
initial increase of the order parameter $m(t)\propto t^{\theta^{\prime}}$,
respectively. The critical exponents $\theta$ and $\theta^{\prime}$ depend on
the dynamic universality class, \cite{hohenberg-77} and have been calculated for
a number of dynamic models such as the model with a nonconserved order parameter
\cite{janssen-89} (model A), the model with an order parameter coupled to a
conserved density\cite{oerding-93}  (model C), and the models with reversible mode
coupling\cite{oerding-93-2} (models E, F, G, and J). The universal scaling behavior
of the initial stage of the critical relaxation has been verified by extensive
numerical simulations.\cite{zheng-98}

The influence of various kinds of disorder on phase transitions is one of the
central problems in condensed-matter physics.
\cite{stichcombe-83,chardy-96,pelissetto-02,alonso-01,holovatch-02,grinstein-76}
In the present paper, we focus our attention on the nonequilibrium
critical properties of systems in the presence of quenched temperaturelike
disorder (for example, magnetic systems with quenched nonmagnetic impurities).
In the simplest case, the disorder may be viewed as randomly distributed
pointlike defects. \cite{grinstein-76} According to the  Harris
criterion,\cite{harris-74} the quenched uncorrelated pointlike defects
change the critical behavior only if the heat capacity exponent of the
pure system is positive $\alpha_p>0$. As a result, the pointlike disorder is
relevant only for Ising-like three-dimensional ($d=3$) systems.
The short-time critical behavior of systems with pointlike uncorrelated
defects was studied in Refs.~\onlinecite{kissner-92} and \onlinecite{oerding-95}.
Real systems, however, often contain defects in the form of linear
dislocation, planar grain boundaries, three-dimensional cavities, or other
extended defects. One possibility to treat the systems with extended
defects is the model suggested by Weinrib and Halperin (WH), \cite{weinrib-83}
in which defects are long-range correlated and characterized by
a correlation function that has a power-law decay $g(x) \propto x^{-a}$
with distance $x$. This type of disorder has a direct interpretation for
integer values $a$: the case $a=d$ corresponds to uncorrelated pointlike
defects, while $a=d-1$ ($a=d-2$) describes infinite lines (planes) of
defects of random orientation. The statics of the WH model and its dynamics
not far from equilibrium were examined  by means of the renormalization-group
(RG) methods to two-loop order using both the double  $\varepsilon=4-d$,
$\delta=4-a$  expansion\cite{korucheva-98} and the direct calculations in
$d=3$.\cite{fedorenko-00} The nonequilibrium critical relaxation of the WH
model  was considered only in the one-loop approximation, \cite{chen-01}
although numerous investigations of pure and disordered systems performed
with the use of the field-theoretic approach show that the predictions made in
the one-loop approximation, especially on the basis of the $\varepsilon$
expansion, can differ from the real critical behavior.
\cite{holovatch-02,fedorenko-00}
The possible alternative to the isotropic scenario realized in
the WH model is the anisotropic scenario realized in the
model proposed by Dorogovtsev, \cite{dorogovtsev-80}
in which defects are strongly correlated in $\varepsilon_d$
dimensions and randomly distributed over the remaining $\tilde{d}%
=d-\varepsilon_d$ dimensions. The case $\varepsilon_d=0$ is associated with
uncorrelated pointlike  defects, while the extended linear (planar) defects
are related to the cases $\varepsilon_d=1$(2). The case of the noninteger value
of $\varepsilon_d$  may be related to a system containing fractal-like
defects so that $\varepsilon_d$ is  interpreted as an effective fractal dimension
of a complex random defect system.\cite{yamazaki-88}
The statics and equilibrium dynamics of Dorogovtsev's model were studied
in Refs.~\onlinecite{boyanovsky-82,prudnikov-83,lawrie-84,blavatska-03}
(for a review of the models with extended defects, see
Refs.~\onlinecite{decesare-94,korzhenevskii-94,yamazaki-86}).

In the present paper, we investigate the short-time critical dynamics
of Dorogovtsev's model with a nonconserved order parameter
using a field-theoretic approach in the two-loop approximation.
The paper is organized as follows. Section \ref{sec2} introduces the model
describing the critical dynamics of disordered systems with
extended defects.
In Sec.~\ref{sec3}, the corresponding effective field theory is renormalized
up to two-loop order.
In Sec.~\ref{sec4}, we derive the asymptotic behavior of response and correlation
functions and calculate the critical exponents using the Pad\'{e}-Borel
resummation method. The final section contains our conclusions.

%%%%%%%%%%%%%%%%%%%%%%%
\section{The model}
%%%%%%%%%%%%%%%%%%%%%%%
\label{sec2}

In equilibrium at temperature $T$ the Hamiltonian describing the disordered
systems with extended defects is given by\cite{dorogovtsev-80}
\begin{eqnarray}
&&\mathcal{H}_{V}[\phi ]=\int d^{d}x\left[ \frac{1}{2}\sum_{\beta
=1}^{n}[\tau(\phi ^{\beta })^{2}+\alpha^{2}|\nabla _{\parallel }\phi ^{\beta
}|^{2}\right.  \notag \\
&&+|\nabla _{\perp }\phi ^{\beta }|^{2}+\left. V(\mathbf{x}_{\perp})(\phi
^{\beta })^{2}]+\frac{g\alpha^{\varepsilon _{d}}}{4!}(\sum_{\beta
=1}^{n}(\phi ^{\beta })^{2})^{2}\right] ,  \label{eq:H}
\end{eqnarray}%
where $\phi ^{\beta }(\mathbf{x})$ is the $n$-component order parameter, $%
\tau$ is the reduced temperature, and $g$ is a positive constant.
$\alpha$ is the bare anisotropy constant and the factor
$\alpha^{\varepsilon_d}$ in the quartic term is included for technical
convenience.\cite{lawrie-84}
$V(\mathbf{x_{\perp }})$ is the potential of defects, which can be viewed
as $\varepsilon_d$-dimensional objects, each extending throughout the whole
system along the coordinate $\mathbf{x}_{\parallel}$, while in the
transverse direction $\mathbf{x}_{\perp}$ they are randomly distributed with
the concentration taken to be well below the percolation limit.
The potential $V(\mathbf{x_{\perp }})$ is assumed
to be Gaussian-distributed with the zero mean and the second cumulant
\begin{equation}
\langle V(\mathbf{x_{\perp }})V(\mathbf{x_{\perp}^{\prime}})\rangle
=\Delta \delta ^{d-\varepsilon _{d}}(\mathbf{x_{\perp }}-\mathbf{x_{\perp
}^{\prime}}).
\end{equation}%
The dynamics of the non-conserved order parameter $\phi ^{\beta }(\mathbf{x})$
in the system defined by Hamiltonian (\ref{eq:H}) can be expressed in the form
of the Langevin equation,\cite{hohenberg-77}
\begin{equation}
\frac{\partial {\phi ^{\beta }(\mathbf{x},t)}}{\partial t}=-\lambda \frac{%
\delta \mathcal{H}_{V}[\phi ]}{\delta {\phi ^{\beta }(\mathbf{x},t)}}
+\eta^{\beta }(\mathbf{x},t),  \label{eq:Langevin}
\end{equation}%
where $\lambda$ is the Onsager kinetic coefficient, and the function $%
\eta ^{\beta }(\mathbf{x},t)$ is a Gaussian random-noise source with
correlations
\begin{equation}
\overline{\eta ^{\beta }(\mathbf{x},t)\eta ^{\gamma }(\mathbf{x^{\prime}}%
,t^{\prime})}=2\lambda \delta _{ \beta \gamma }\delta (\mathbf{x}-%
\mathbf{x^{\prime}})\delta (t-t^{\prime}). \label{noise}
\end{equation}%
Here, we summarize the main features of model (\ref{eq:H})-(\ref{noise})
revealed in Refs.~\onlinecite{dorogovtsev-80} and
\onlinecite{boyanovsky-82,lawrie-84,prudnikov-83,blavatska-03}.
The Harris criterion is modified in the presence of extended
defects.\cite{boyanovsky-82} The disorder affects the critical behavior only
if the corresponding crossover exponent $\varphi$ is positive,
\begin{equation}
\varphi=\alpha_p+\nu_p\varepsilon_d>0, \label{harris}
\end{equation}
where $\nu_p$ is the correlation length exponent of the pure system.
Consequently, the disorder with extended defects is relevant for $d=3$ over
a wider range of $n$ than the pointlike  disorder.\cite{blavatska-03}
The critical properties of the  model can be examined  within the
RG framework  by using the double expansion in both $\varepsilon$ and $\varepsilon_d$,
which was suggested in Ref.~\onlinecite{dorogovtsev-80}.
Due to the spatial anisotropy caused by disorder, two correlation
lengths $\xi_{\perp}$ and $\xi_{\parallel}$ naturally arise, one of
which  is perpendicular to the extended defects direction whereas another
is parallel to this direction. In the critical point, their divergences
are characterized by corresponding critical exponents $\nu_{\perp}$ and
$\nu_{\parallel}$ :  $\xi_{\perp}\propto |\tau|^{-\nu_{\perp}}$,
$\xi_{\parallel}\propto |\tau|^{-\nu_{\parallel}}$. The correlation
of the order-parameter fluctuations in two different points depends on the
orientation of their distance vector, so that the behavior of the correlation function
is characterized by a pair of Fisher exponents $\eta_{\perp}$ and $\eta_{\parallel}$.
As was shown in Ref.~\onlinecite{prudnikov-83}, the equilibrium dynamics is also
modified and there are two dynamic exponents $z_{\perp}$ and $z_{\parallel}$.
On the other hand, as the interaction of all order-parameter components with disorder
is the same, the susceptibility (as well as the order parameter and heat capacity)
is characterized by the single exponent $\gamma$ and can be written as
\cite{boyanovsky-82,prudnikov-83,lawrie-84,blavatska-03}
\begin{equation}
\chi(q_{\perp},q_{\parallel},t,\tau)=\tau^{-\gamma}{\cal G}_1(\tau^{-\nu_{\perp}}q_{\perp},
\tau^{-\nu_{\parallel}}q_{\parallel},t\tau^{\nu_{\perp}z_{\perp}}),
\end{equation}
or in the critical point ($\tau\to 0$) as
\begin{eqnarray}
 \chi(q_{\perp},q_{\parallel},t,0)&=&|q_{\perp}|^{\eta_{\perp}-2}{\cal G}_2
(q_{\parallel}|q_{\perp}|^{-\nu_{\perp}/\nu_{\parallel}},
t|q_{\perp}|^{z_{\perp}}) \notag \\
&=&|q_{\parallel}|^{\eta_{\parallel}-2}{\cal G}_3
(q_{\perp}|q_{\parallel}|^{-\nu_{\parallel}/\nu_{\perp}},
t|q_{\parallel}|^{z_{\parallel}}),
\end{eqnarray}
where $q_{\perp}$, $q_{\parallel}$ are the components of the momenta along
the $\mathbf{x}_{\perp}$ and $\mathbf{x}_{\parallel}$ directions, respectively,
and ${\cal G}_i$ are the scaling functions. The  critical exponents defined
above were calculated to second order in $\varepsilon$ and  $\varepsilon_d$ in
Refs.~\onlinecite{prudnikov-83} and \onlinecite{lawrie-84}.

It has been recently argued in Ref.~\onlinecite{vojta-03} that the
planar ($\varepsilon_d=2$) defects can destroy the sharp continuous transition in
the Ising-like systems due to the existence of rare infinite spatial regions which
are devoid of defects and therefore may be locally in the ordered phase.
It had been proposed early that the similar effects can give rise to the
instability of the usual critical behavior of disordered systems
with respect to the replica symmetry breaking \cite{dotsenko-95} (RSB)
and even lead to the appearance of an intermediate spin-glass phase.\cite{dotsenko-02}
However, for pointlike defects the rare regions have finite size and thus cannot
develop the true static order, so that the order parameter on such rare regions
still fluctuates. The more accurate RG calculations performed in
Ref.~\onlinecite{fedorenko-01} have shown the stability of the critical behavior of
the weakly disordered systems with respect to RSB potentials. In the
case of extended defects, these regions are infinite in $\varepsilon_d$ dimensions
and, according  to the Mermin-Wagner theorem,\cite{zinn-justin} can develop the
true static long-range order for $n=1$ and $\varepsilon_d\geq 2$ [in two-dimensional
systems with continuous symmetry ($n>1$) there is no true static long-range order].
As suggested in Ref.~\onlinecite{vojta-03}, the last can result in the smearing of
the phase transition. Although this picture, based on the extremal statistics, is
supported by the accompanying numerical simulations,\cite{vojta-03} there is a
contradiction with the early numerical studies.\cite{lee-92} Considering in what
follows the particular case $n=1$ and $\varepsilon_d=2$,
we will not take into account these effects.

We now consider the relaxation of model (\ref{eq:H})-(\ref{noise})
from a nonequilibrium state with a small correlation length. This initial
state can be macroscopically prepared by quenching the system from
a high temperature $T_{0}\gg T_{c}$ to the critical temperature $T_c$.
We have to specify the distribution of the initial condition
$\phi _{0}^{\beta }(\mathbf{x})=\phi^{\beta }(\mathbf{x},t=0)$.
The reasonable assumption is that the distribution is Gaussian,
\begin{equation}
P[\phi _{0}]\propto \exp \left(- \frac{\tau _{0}}{2}\int d^{d}x
(\phi _{0}^{\beta }(\mathbf{x})-m_{0}^{\beta }(\mathbf{x}))^{2}\right),
\label{H-i}
\end{equation}%
where $m^{\beta}_{0}(\mathbf{x})$ is the initial order parameter and
$\tau _{0}^{-1}$ is the width of the initial distribution.
Equation (\ref{H-i}) guarantees that the initial correlations are short-range.
Terms of  higher order in the exponent of Eq.~(\ref{H-i}) turn out to be
irrelevant in RG sense. By introducing a Martin-Siggia-Rose response
field\cite{martin-73} $\tilde{\phi}^{\beta }(\mathbf{x},t)$,
we can average over the random noise $\eta ^{\beta }(\mathbf{x},t)$
and obtain an equivalent formulation of dynamics in terms of a
generating functional for all response and correlation functions,
\begin{eqnarray}
W[h,\tilde{h}] & =& \ln \int {\cal D} \phi {\cal D} i \tilde{\phi} P[\phi _{0}]
 \exp\left\{-\mathcal{L}_{V}[\phi ,\tilde{\phi}] \right. \notag \\
  & + & \left. \int\limits_{0}^{\infty }dt\int
d^{d}x (h^{\beta }\phi^{\beta }+\tilde{h}^{\beta }\tilde{\phi}^{\beta }) \right\},
\label{gen-fun}
\end{eqnarray}
where the action $\mathcal{L}_{V}[\phi ,\tilde{\phi}]$ is given by
\begin{equation}
\mathcal{L}_{V}[\phi ,\tilde{\phi}]=\int\limits_{0}^{\infty }dt\int d^{d}x\,%
\tilde{\phi}^{\beta }\left[ \frac{\partial {\phi ^{\beta }}}{\partial t}%
+\lambda \frac{\delta \mathcal{H}_{V}[\phi ]}{\delta {\phi ^{\beta }}}%
-\lambda \tilde{\phi}^{\beta }\right] .  \label{eq:L-V}
\end{equation}%
We can perform the average over disorder directly in Eq.~(\ref{gen-fun})
without using the replica trick\cite{de-dominicis-86} and obtain the $V$%
-independent and translational-invariant (however, in space $\mathbf{x}$
only) effective action
\begin{eqnarray}
&& \mathcal{L}[\phi ,\tilde{\phi}] = \int\limits_{0}^{\infty }dt\int
d^{d}x\, \tilde{\phi}^{\beta }\left[ \frac{\partial {\phi ^{\beta }}}{%
\partial t} +\lambda \left(\tau-\alpha^{2}\nabla _{\parallel }^{2} \right.
\right.  \notag \\
&& -\left. \nabla _{\perp }^{2}\right) \phi ^{\beta }+\lambda \frac{%
g\alpha^{\varepsilon _{d}}}{3!}\phi ^{\gamma }\phi ^{\gamma }\phi ^{\beta
}-\lambda \tilde{\phi}^{\beta } - \frac{1}{2}\lambda ^{2}\Delta
\int\limits_{0}^{\infty }dt^{\prime}  \notag \\
&& \left. \times \int d^{d}x^{\prime}\, \delta^{\tilde{d}} ({\mathbf{x}_{\perp }}-{%
\mathbf{x}_{\perp }^{\prime}}) \phi ^{\beta }(\mathbf{x},t) \tilde{\phi}^{\gamma
}(\mathbf{x}^{\prime},t^{\prime})\phi ^{\gamma }(\mathbf{x}^{\prime},t^{\prime}) \right].\ \
\label{eq:L}
\end{eqnarray}%
%%%%%%%%%%%%%%%%%%%%%%%%%%%%%%%%%%%%%%%%%%%%%%%%%%%%%%%%%%%%%%%%%%%%%%%%%%%%%
\begin{figure}[tbph]
\includegraphics[clip,width=3.0in]{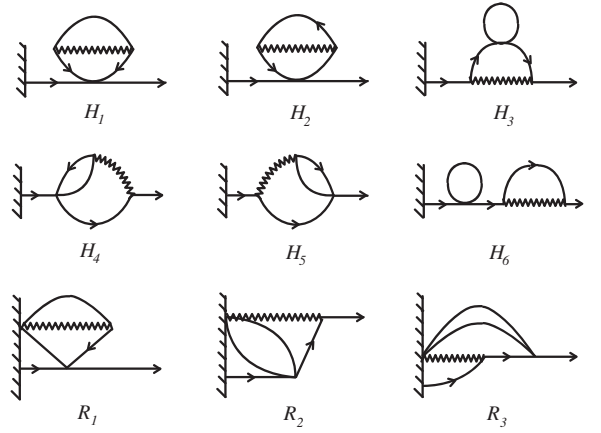}
\caption{Diagrams contributing to $\Gamma_{0,1}(q,t)_{[\tilde{\phi}_0]}$
at two-loop order. Diagrams $H_1-H_6$ contribute to
$\Gamma^{\mathrm{(i)}}_{0,1}(q,t)_{[\tilde{\phi}_0]}$; diagrams $R_1$,$R_2$,
and $R_3$ contribute to $K(q;t^{\prime},t^{\prime\prime})$. The solid lines
correspond to the initial correlator $C^{(i)}_q(t,t')$ in diagrams
$H_1-H_6$ and to the equilibrium correlator $C^{(e)}_q(t,t')$ in $R_1$,$R_2$,
and $R_3$; the lines with an arrow correspond to the propagator $G_q(t,t')$;
the four-leg vertex $\times$ corresponds to the $g$ interaction; the wavy line
corresponds to the disorder interaction $\Delta$;
the wall corresponds to the "time surface" $t=0$.
}
\label{fig1}
\end{figure}
%%%%%%%%%%%%%%%%%%%%%%%%%%%%%%%%%%%%%%%%%%%%%%%%%%%%%%%%%%%%%%%%%%%%%%%%%%%%%
For $g=\Delta =0$, the generating functional (\ref{gen-fun}) becomes Gaussian
and can be easily evaluated in momentum space by solving the corresponding variational
equations. We have to take into account the initial condition (\ref{H-i})
by imposing the boundary conditions $\tilde{\phi}^{\beta}(\mathbf{x},\infty)=0$ and
$\phi_0^{\beta}({\bf x})=m_0^{\beta}({\bf x})+\tau_0^{-1}\tilde{\phi}^{\beta}({\bf x},0)$.
The free response function (propagator) $G_{q}(t,t^{\prime })$ and the
free correlator $C_{q}(t,t^{\prime })$ are then given by
\begin{eqnarray}
&& \mbox{}\hspace{-7mm} G_{q}(t,t^{\prime })=\Theta (t-t^{\prime })
\exp (-\lambda (\tau +\alpha^{2}q_{\parallel }^{2}+q_{\perp }^{2})
(t-t^{\prime })), \\
&& \mbox{}\hspace{-7mm} C_{q}(t,t^{\prime })=C_{q}^{(\mathrm{e)}}
(t,t^{\prime })+C_{q}^{(\mathrm{i)}}(t,t^{\prime }),
\end{eqnarray}
where
\begin{eqnarray}
C_{q}^{(\mathrm{e)}}(t,t^{\prime })& = & \frac{\exp (-\lambda (\tau
+\alpha^{2}q_{\parallel }^{2}+q_{\perp }^{2})|t-t^{\prime }|)}{\tau
+\alpha^{2}q_{\parallel }^{2}+q_{\perp }^{2}}, \label{equilcor}\\
C_{q}^{(\mathrm{i)}}(t,t^{\prime }) &=&(\tau _{0}^{-1}-\frac{1}{\tau
+\alpha^{2}q_{\parallel }^{2}+q_{\perp }^{2}})  \notag \\
&\times &\exp (-\lambda (\tau +\alpha^{2}q_{\parallel }^{2}+q_{\perp
}^{2})(t+t^{\prime }))  \label{initcor}
\end{eqnarray}%
are the equilibrium and the initial nonequilibrium parts of the
Gaussian correlator, respectively.

%%%%%%%%%%%%%%%%%%%%%%%%%%%
\section{Renormalization}
%%%%%%%%%%%%%%%%%%%%%%%%%%%
\label{sec3}

We now exploit the methods of renormalized field theory in conjunction with
a generalized expansion in $\varepsilon $ to investigate the scaling
behavior of nonequilibrium response and correlation functions. We define the
full one-particle reducible Green functions as
\begin{equation}
G^{\tilde{L}}_{\tilde{N},N} := \overline{\langle [\phi]^{N}
[\tilde{\phi}]^{\tilde{N}} [\tilde{\phi}_0]^{\tilde{L}} \rangle}.
\label{green-function}
\end{equation}
The Green functions $G^{\tilde{L}}_{\tilde{N},N}$ may be calculated by a
perturbation expansion in the coupling constants $g$ and $\Delta$. The
Feynman diagrams that contribute to the Green functions computed using the
effective action (\ref{eq:L}) involve momentum integrations of dimensions
$d=4-\varepsilon$ and $\tilde{d}=4-\tilde{\varepsilon }$. We use the dimensional
regularization\cite{zinn-justin} to calculate integrals and employ the minimal
subtraction scheme\cite{zinn-justin} to absorb the remaining poles
in $(c_1\varepsilon+c_2\tilde{\varepsilon})$ into multiplicative Z factors and to
introduce the renormalized quantities according to
\begin{eqnarray}
&& \phi_b = Z_s^{1/2}\phi,\ \ \tilde{\phi}_b = Z_{\tilde{s}}^{1/2}\tilde{\phi},
  \ \ \tilde{\phi}_{0b}=(Z_{\tilde{s}} Z_0)^{1/2}\tilde{\phi}_0,  \notag \\
&& \tau_b=(Z_{\tau}/Z_s)\tau,\ \ \
\lambda_b=(Z_s/Z_{\tilde{s}})^{1/2}\lambda, \ \ \alpha_b=Z_{\alpha} \alpha,
 \ \ \ \label{rencond}  \\
&&  g_b=Z_s^{-2}Z_{g}g\mu^{\varepsilon}, \ \
 \Delta_b=Z_s^{-2}Z_{\Delta}\Delta \mu^{\tilde{\varepsilon}}, \notag
\end{eqnarray}%
where the subscript 'b' denotes bare quantities and $\mu$ is an external momentum scale.
Naive dimensional analysis gives $\tau_0\propto \mu^2$ and, thus, $\tau_0^{-1}$ is
an irrelevant parameter in the RG sense. It is convenient to consider the Dirichlet
boundary conditions $\tau_0^{-1}=0$ and $m_0^{\beta}(\mathbf{x})=0$. The general
case is recovered by treating the parameters $\tau_0^{-1}$ and
$m_0^{\beta}(\mathbf{x})$ as additional perturbations.
The Z factors, except for $Z_0$,
were calculated in Ref.~\onlinecite{lawrie-84} to two-loop order. The new
factor $Z_0$ serves to cancel the divergences arising from the initial part $%
C_{q}^{(\mathrm{i})}(t,t^{\prime})$ for $t+t^{\prime}\to 0$. Note that although
there exist two different Z factors for fields $\phi$ and $\tilde{\phi}$, we need
only $Z_0$ to renormalize fields $\phi_0$ and $\tilde{\phi}_0$. This is due to
the Ward identities
\begin{eqnarray}
&& \phi^{\beta}_0(\mathbf{x})= \tau_0^{-1}\tilde{\phi}^{\beta}_0(\mathbf{x}), \\
&& \partial_t \phi^{\beta}(\mathbf{x},t)|_{t=0}\equiv \dot{\phi}^{\beta}_0(\mathbf{x})
 =2\lambda\tilde{\phi}^{\beta}_0(\mathbf{x}), \label{relation}
\end{eqnarray}
which hold when inserted in the connected Green functions.\cite{janssen-89}
In order to determine $Z_0$, we require that
\begin{equation}
(Z_0Z_sZ_{\tilde{s}})^{-1/2}G^{1}_{0,1}(\mathbf{x},t) = \mathrm{\ finite \ \
for \ } \varepsilon,\ \tilde{\varepsilon} \to 0. \label{rencond2}
\end{equation}
%%%%%%%%%%%%%%%%%%%%%%%%%%%%%%%%%%%%%%%%%%%%%%%%%%%%%%%%%%%%%%%%%%%%%%%%%%%%%%%%%%%%
\begin{table}[tbp]
\caption{Values of two-loop diagrams contributing to
$\Gamma_{0,1}(q=0,t)_{[\tilde{\phi}_0]}$.
The common weight factor for all diagrams is $(n+2)/12 (K_dg)(K_{\tilde{d}%
}\Delta)\lambda\Gamma(1-(\protect\varepsilon+\tilde{\protect\varepsilon})/2) (2%
\protect\lambda t)^{-1+(\protect\varepsilon+\tilde{\protect\varepsilon})/2}$.}
\label{table1}%
\begin{ruledtabular}
\begin{tabular}{ll}
$H_1$  &   $-1/\tilde{\varepsilon}-3/4-\varepsilon/4\tilde{\varepsilon}$ \\
$H_2$  &   $1/\tilde{\varepsilon} $ \\
$H_3$  &   $1/\tilde{\varepsilon}-\ln 2/ 2 $ \\
$H_4$  &   $2/\tilde{\varepsilon}+1-2\ln 2 $ \\
$H_5$  &   $2/\tilde{\varepsilon}+1+5\ln 2 -3\ln 3 -\sqrt{3}\ln(2+\sqrt{3})$ \\
$H_6$  &   $-1/\tilde{\varepsilon}+\ln 2/ 2 $ \\ \hline
$R_1$  &   $2/\tilde{\varepsilon}$ \\
$R_2$  &   $2 \ln2 $ \\
$R_3$  &   $3\ln 3 - 4 \ln2 $ \\
\end{tabular}
\end{ruledtabular}
\end{table}
%%%%%%%%%%%%%%%%%%%%%%%%%%%%%%%%%%%%%%%%%%%%%%%%%%%%%%%%%%%%%%%%%%%%%%%%%%%%%%%%%%%%
It is more convenient to work with the Fourier transform $%
G^{1}_{0,1}(q,t)=\int d^d x\exp(-{i\mathbf{q x}})G^{1}_{0,1}(\mathbf{x},t)$,
which can be written in the following form:
\begin{equation}
G^{1}_{0,1}(q,t) = \int\limits_0^t dt^{\prime}\bar{G}_{1,1}(q;t,t^{\prime})
\Gamma^{\mathrm{(i)}}_{0,1}(q,t^{\prime})_{[\tilde{\phi}_0]},  \label{G101}
\end{equation}
where $\bar{G}_{1,1}(q;t,t^{\prime})$ denotes the full unrenormalized
two-point function coming only from the equilibrium part of the correlator
(\ref{equilcor}). $\Gamma^{\mathrm{(i)}}_{0,1}(q,t^{\prime})_{[\tilde{\phi}_0]}$
is a reducible one-point vertex function with a single insertion of the
response field $\tilde{\phi}_0$, which contains in its last irreducible part
at least one initial correlator (\ref{initcor}). Figure~1 shows the two-loop
diagrams $H_1$ -- $H_6$ which contribute to
$\Gamma^{\mathrm{(i)}}_{0,1}(q,t^{\prime})_{[\tilde{\phi}_0]}$
together with those computed for pure systems in Ref.~\onlinecite{janssen-89}.
The  singular parts of computed diagrams are given in Table~\ref{table1}.
Note that although $\bar{G}_{1,1}(q;t,t^{\prime})$ is calculated with
equilibrium propagators and
correlators, it is different from the translational-invariant equilibrium
response function $G^{\mathrm{(eq)}}_{1,1}(q;t-t^{\prime})$ because of the
restriction of internal time integration to positive times $t\ge 0$. Indeed,
the equilibrium Green functions must be calculated by means of the effective
action (\ref{eq:L}) with the lower limit of time integration $t=0$
replaced by $t=-\infty$. We can relate $\bar{G}_{1,1}(q;t,t^{\prime})$ to $%
G^{\mathrm{(eq)}}_{1,1}(q,t-t^{\prime})$ using the method suggested in Ref.~%
\onlinecite{oerding-93}. Following Ref.~\onlinecite{oerding-93}, we choose
the equilibrium distribution
$P_{\mathrm{eq}}[\phi_0]\propto\exp(-{\cal H}_V[\phi_0])$ instead
of distribution (\ref{H-i}) to average over the initial field $\phi_0$ in
Eq.~(\ref{gen-fun}).
This allows us to obtain a perturbation series for $G^{\mathrm{(eq)}%
}_{1,1}(q,t-t^{\prime})$ where internal time integrations range from zero to
infinity. However, the non-Gaussian probability distribution $P_{\mathrm{eq}}[\phi_0]$
generates in the effective action (\ref{eq:L}) an additional vertex,
\begin{eqnarray}
&-&\frac12 \lambda {\Delta} \int\limits_{0}^{\infty }dt \int d^{d}x\,
\int d^{d}x^{\prime}\, \delta^{\tilde{d}} ({\mathbf{x}_{\perp }}
  -{\mathbf{x}_{\perp }^{\prime }}) \notag \\
&\times &\tilde{\phi} ^{\beta }(\mathbf{x},t) \phi ^{\beta }(\mathbf{x},t)
\phi_0 ^{\gamma }(\mathbf{x}^{\prime}) \phi_0 ^{\gamma }
 (\mathbf{x}^{\prime}), \label{newvertex}
\end{eqnarray}
which is located at "time surface" $t=0$. We can relate
$G^{\mathrm{(eq)}}_{1,1}(q,t-t^{\prime})$ to $\bar{G}_{1,1}(q;t,t^{\prime})$
by using the following equation:
\begin{equation}
G^{\mathrm{(eq)}}_{1,1}(q,t-t^{\prime}) = \int\limits_{t^{\prime}}^t
dt^{\prime\prime}\bar{G}_{1,1}(q;t,t^{\prime\prime}) \Gamma^{\mathrm{(eq)}%
}_{0,1}(q,t^{\prime\prime})_{[\tilde{\phi}(t^{\prime})]},  \label{Geq}
\end{equation}
where $\Gamma^{\mathrm{(eq)}}_{0,1}(q,t^{\prime\prime})_{[\tilde{\phi}%
(t^{\prime})]}$ is a reducible vertex function with a single insertion of $%
\tilde{\phi}(t^{\prime})$, which is calculated with equilibrium correlators
and contains in its last irreducible part at least one of the new vertex
(\ref{newvertex}).
In contrast to the pure system, \cite{janssen-89} the disordered system
(\ref{eq:H})-(\ref{noise}) has already the nontrivial contributions to
$\Gamma^{\mathrm{(eq)}}_{0,1}(q,t^{\prime\prime})_{[\tilde{\phi}(t^{\prime})]}$
at two-loop level. The corresponding two-loop diagrams $R_1$ -- $R_3$
are given in Fig.~1.
Considering Eq.~(\ref{Geq}) as an integral transformation with kernel $\Gamma^{%
\mathrm{(eq)}}_{0,1}(q,t^{\prime\prime})_{[\tilde{\phi}(t^{\prime})]}$, we
can find the kernel of the inverse transformation $K(q;t,t^{\prime})$ order
by order in perturbation theory using the following integral equation:
\begin{equation}
\int\limits_{t^{\prime}}^t dt^{\prime\prime}\Gamma^{\mathrm{(eq)}%
}_{0,1}(q,t)_{[\tilde{\phi}(t^{\prime\prime})]}
K(q;t^{\prime\prime},t^{\prime})=\delta(t-t^{\prime}). \label{eq18}
\end{equation}
Combining Eqs.~(\ref{G101}), (\ref{Geq}), and (\ref{eq18}), we arrive at
\begin{equation}
G^{1}_{0,1}(q,t) = \int\limits_{0}^t dt^{\prime}G^{\mathrm{(eq)}%
}_{1,1}(q,t-t^{\prime}) \Gamma_{0,1}(q,t^{\prime})_{[\tilde{\phi}_0]},
\label{eq19}
\end{equation}
where
\begin{equation}
\Gamma_{0,1}(q,t^{\prime})_{[\tilde{\phi}_0]} = \int\limits_{0}^{t^{\prime}}
dt^{\prime\prime}K(q;t^{\prime},t^{\prime\prime}) \Gamma^{\mathrm{(i)}%
}_{0,1}(q,t^{\prime\prime})_{[\tilde{\phi}_0]}. \label{vertex-full}
\end{equation}
The two-loop result for the singular part of  vertex function
(\ref{vertex-full}) is given by
\begin{widetext}
\begin{eqnarray}
\Gamma_{0,1}(q=0,t)_{[\tilde{\phi}_0]}^{\mathrm{sing}}  &=& \delta(t) +
\frac{\left( n+2\right) }{12}\lambda g\Gamma \left( 1-\,\frac{%
\varepsilon }{2}\right)
 \left( 2\lambda t\right) ^{-1+\varepsilon /2}\notag\\
&-&\frac{%
\left( n+2\right) ^{2}}{36\varepsilon }\lambda {g}^{2}\Gamma ^{2}\left( 1-%
\frac{\varepsilon }{2}\right)
 \left( {\frac{\Gamma ^{2}\left(
1+\varepsilon/2 \right) }{\Gamma \left( 1+\varepsilon \right) }}-\frac{1}{2}%
\right) \left( 2\lambda t\right) ^{-1+\varepsilon } \notag\\
&+& \,\frac{\left( n+2\right) }{12}\lambda {g}^{2}\Gamma \left( 1-\varepsilon
\right) \left( -\frac{1}{\varepsilon }+\ln 2-\frac{1}{2}\right) \left(
2\lambda t\right) ^{-1+\varepsilon } \notag\\
&+&\frac{\left( n+2\right) ^{2}}{%
144\varepsilon }\,\lambda {g}^{2}\Gamma ^{2}\left( 1-\frac{\varepsilon }{2}%
\right) \left( 2\lambda t\right) ^{-1+\varepsilon } \notag\\
&+&\frac{\left( n+2\right) }{48}\lambda g\Delta
\Gamma \left( 1-\,\frac{%
\varepsilon +\tilde{\varepsilon }}{2}\right) \left( \frac{24}{\tilde{%
\varepsilon }}+A-2-\,{\frac{\varepsilon }{\tilde{\varepsilon }}}\right)
\left( 2\lambda t\right) ^{-1+(\varepsilon +\tilde{\varepsilon })/2},
\label{Gamma10}
\end{eqnarray}
\end{widetext}
where $A=7+4\ln 2-4\sqrt{3}\ln(2+\sqrt{3})$, $\Gamma(x)$ is the
Euler gamma function, and we have
absorbed factors of $K_d=2\pi^{d/2}/((2\pi)^d\Gamma(d/2))$
and $K_{\tilde{d}}$ into the redefinition of
the coupling constants $g$ and $\Delta$, respectively.
Taking into account Eq.~(\ref{eq19}), the renormalization condition (\ref{rencond2})
can be rewritten in the form \cite{janssen-89,oerding-93}
\begin{equation}
Z_0^{-1/2} \int\limits_0^{\infty} dt e^{-i\omega t} \Gamma_{0,1}(0,t)_{[%
\tilde{\phi}_0]}^{\mathrm{sing}} = \mathrm{\ finite \ \ for \ } \varepsilon,\
\tilde{\varepsilon} \to 0,
\end{equation}
where $\Gamma_{0,1}(0,t)_{[\tilde{\phi}_0]}^{\mathrm{sing}}$ is given by
Eq.~(\ref{Gamma10}) expressed in terms of the renormalized quantities
using Eqs.~(\ref{rencond}). They read in explicit form
\begin{eqnarray}
&& g \to g_b=g\left(1+\frac{(n+8)g}{6\varepsilon}-\frac{6\Delta}{\tilde{\varepsilon}}
 +O({\varepsilon}^2,{\varepsilon}\tilde{\varepsilon},\tilde{\varepsilon}^2)\right),
  \notag \\
&& \lambda \to \lambda_b=\lambda\left(1+\frac{\Delta}{\tilde{\varepsilon}}+
O({\varepsilon}^2,{\varepsilon}\tilde{\varepsilon},\tilde{\varepsilon}^2)\right).
\notag
\end{eqnarray}
As a result, at two-loop order we obtain
\begin{eqnarray}
Z_0 & = & 1+ \frac{n+2}{6\varepsilon}g+\frac{n+2}{12\varepsilon^2} \left[
\frac{n+5}{3}+\left(\ln 2 -\frac12\right)\varepsilon \right] g^2  \notag \\
& + & \frac{n+2}{24}\frac{(\varepsilon^2-24\tilde{\varepsilon}+ A\varepsilon%
\tilde{\varepsilon})}{\varepsilon\tilde{\varepsilon}(\varepsilon +\tilde{%
\varepsilon})}g\Delta. \label{Z0}
\end{eqnarray}
For $n=1$ and $\tilde{\varepsilon}=\varepsilon$, Hamiltonian (\ref{eq:H})
corresponds to the Ising model with quenched pointlike disorder, so that
Eq.~(\ref{Z0}) reduces to $Z_0$ obtained in Ref.~\onlinecite{oerding-95}.

%%%%%%%%%%%%%%%%%%%%%%%%%%%%%%%%%%%%%%%%%%%%%
\section{Scaling and critical exponents}
%%%%%%%%%%%%%%%%%%%%%%%%%%%%%%%%%%%%%%%%%%%%%
\label{sec4}

We are now in a position to discuss the scaling properties of Green
functions (\ref{green-function}). Using the fact that the bare Green
function $G^{\tilde{L}}_{\tilde{N},Nb}$ does not depend
on the external momentum scale $\mu$ introduced in Eqs.~(\ref{rencond}), we
can derive the RG equation in the usual way. \cite{zinn-justin} It reads
\begin{eqnarray}
\left[ \hat{R}  +\frac{N}2\gamma+\frac{\tilde{N}}2 \tilde{\gamma}
+ \frac{\tilde{L}}2(\tilde{\gamma}+\gamma_0)\right]G^{\tilde{L}}_{\tilde{N},N} = 0,
\label{rg-eq}
\end{eqnarray}
where we have introduced the operator $\hat{R} \equiv \mu {\partial}_{\mu}+%
\beta_g{\partial}_g+\beta_{\Delta}{\partial}_{\Delta}+
\gamma_{\tau}\tau\partial_{\tau}+\gamma_{\lambda}\lambda\partial_{\lambda}+%
\gamma_{\alpha}\alpha\partial_{\alpha}$, and the $\beta$ and $\gamma$ functions
are defined as derivatives at constant bare parameters,
\begin{eqnarray}
&&\beta_{g}=\left. \mu {\partial}_{\mu} \ln g \right|_{0}, \
\beta_{\Delta}=\left. \mu {\partial}_{\mu} \ln \Delta \right|_{0}, \
\gamma_{\alpha}=\left. \mu {\partial}_{\mu} \ln \alpha \right|_{0}, \  \notag \\
&& \gamma_{\lambda}=\left. \mu {\partial}_{\mu} \ln \lambda \right|_{0} =
 (\tilde{\gamma}-\gamma)/2, \ \ \
\gamma_{\tau}=\left. \mu {\partial}_{\mu} \ln \tau \right|_{0}, \label{gamma-beta} \\
&& \gamma=\left. \mu {\partial}_{\mu} \ln Z_{s} \right|_{0}, \
\tilde{\gamma}=\left. \mu {\partial}_{\mu} \ln Z_{\tilde{s}} \right|_{0}, \
\gamma_0=\left. \mu {\partial}_{\mu} \ln Z_0 \right|_{0}. \ \notag
\end{eqnarray}
Functions (\ref{gamma-beta}), except for $\gamma_0$, were calculated in
Ref.~\onlinecite{lawrie-84} to two-loop order. The new function
$\gamma_0$ reads
\begin{equation}
\gamma_0 = - \frac{n+2}6 g \left[1+ \left(\ln 2 - \frac12 \right) g
+ \frac14 \left(A+\frac{\varepsilon}{\tilde{\varepsilon}}\right)\Delta\right].
\label{gamma0}
\end{equation}
The nature of the critical behavior is determined by the existence of a stable
fixed point (FP) satisfying the simultaneous equations
\begin{equation}
\beta_g(g^*,\Delta^*)=0, \ \ \ \beta_{\Delta}(g^*,\Delta^*)=0. \label{fp-def}
\end{equation}
The full set of FPs exhibited by Eqs.~(\ref{fp-def}) was discussed in
Refs.~\onlinecite{dorogovtsev-80} and
\onlinecite{boyanovsky-82,prudnikov-83,lawrie-84}.
We are interested in the FP
which controls the critical behavior affected by extended defects. The
corresponding FP is given for $n>1$ by \cite{lawrie-84}
\begin{eqnarray}
&&  g^{*}=\frac {3(3\tilde{\varepsilon}-2\varepsilon)}{2(n-1)}+\frac{3}{256}
    \left(-4(31n^2-68n+196){\varepsilon}^{2}\right. \notag \\
&&  +4(73n^2+52n+352)\varepsilon\tilde{\varepsilon}-
    (111n^2+696n+624){\tilde{\varepsilon}}^2 \notag \\
&&  \left. -32{(n+2)(n-1){\varepsilon}^{3}}{\tilde{\varepsilon}}^{-1}\right)
     (n-1)^{-3}, \notag \\
&& \Delta^{*}= {\frac {(n+8)\tilde{\varepsilon}-2(n+2)\varepsilon}{8(n-1)}}
   + O({\varepsilon}^2,{\varepsilon}\tilde{\varepsilon},\tilde{\varepsilon}^2).
   \label{fp}
\end{eqnarray}
In the case $n=1$, there is an accidental degeneracy in Eqs.~(\ref{fp-def}),
which leads to FPs  of order $\varepsilon^{1/2}$ and
$\tilde{\varepsilon}^{1/2}$. The FP corresponding to the systems with extended defects
reads
\begin{equation}
g^{*}=4\Delta^{*}=8[(9\tilde{\varepsilon}
 -6\varepsilon )/106]^{1/2}.  \label{fp-n-1}
\end{equation}

The solution of Eq.~(\ref{rg-eq}) and the simple dimensional analysis yield
the scaling behavior of the Green functions at FPs,
\begin{eqnarray}
&&\mbox{}\hspace{-7mm} G^{\tilde{L}}_{\tilde{N},N}(\{\mathbf{x}_{\parallel},\mathbf{x}_{\perp},t\},
   \tau,\lambda,\alpha,g^*,\Delta^*;\mu)=l^{\varepsilon_d (1-\zeta)
   (N+\tilde{N}+\tilde{L}-2)/2} \notag \\
&& \mbox{}\hspace{-5mm} \times l^{(d-2+\eta_{\perp})N/2+(d+2+\tilde{\eta})\tilde{N}/2
   +(d+2+\eta_0+\tilde{\eta})\tilde{L}/2} \notag \\
&& \mbox{}\hspace{-5mm} \times G^{\tilde{L}}_{\tilde{N},N}(\{\mathbf{x}_{\parallel}l^{\zeta},
   \mathbf{x}_{\perp}l,tl^{z_{\perp}}\},
  \tau l^{-1/\nu_{\perp}},\lambda ,\alpha, g^*,\Delta^*;\mu), \label{scaling}
\end{eqnarray}
where the critical exponents are given by
\begin{eqnarray}
&& \zeta=1-\gamma_{\alpha}(g^*,\Delta^*), \ \
   1/\nu_{\perp}=2-\gamma_{\tau}(g^*,\Delta^*), \ \notag \\
&&\eta_{\perp}=\gamma(g^*,\Delta^*), \ \ \ \ \ \ \
   z_{\perp}=2+\gamma_{\lambda}(g^*,\Delta^*),  \ \\
&&\tilde{\eta}=\tilde{\gamma}(g^*,\Delta^*), \ \ \ \ \ \ \ \ \ \notag
   \eta_{0}=\gamma_0(g^*,\Delta^*).
\end{eqnarray}
The first factor on the right hand side of Eq.~(\ref{scaling}) arises
from the dependence of Green functions on $\alpha$ not only through
$\mathbf{x}_{\parallel}$, but also through an overall factor, which can
be easily found out by inspection of the Feynman diagrams (see the Appendix of
Ref.~\onlinecite{lawrie-84}). Using Eq.~(\ref{scaling}), we reveal
the known relations for the critical exponents,
\cite{dorogovtsev-80,boyanovsky-82,prudnikov-83}
\begin{equation}
\zeta=\nu_{\parallel}/\nu_{\perp}
 =(2-\eta_{\perp})/(2-\eta_{\parallel})=z_{\perp}/z_{\parallel}.
 \label{scal-rel-1}
\end{equation}
Substituting FP (\ref{fp}) into Eq.~(\ref{gamma0}), we obtain the new critical
exponent $\eta_0$ for $n>1$ as a
double expansion in $\varepsilon$ and $\tilde{\varepsilon}$,
\begin{eqnarray}
\eta_0 &=&  (n+2)(n-1)^{-1}\left[0.5\varepsilon-0.75\tilde{\varepsilon} \right.
  + (n-1)^{-2} \notag \\
&\times & \left((0.201599n^2+0.601119n+1.992204)\tilde{\varepsilon}^2 \right. \notag\\
&-& (0.553223n^2-0.400168n+3.573508)\tilde{\varepsilon}\varepsilon \notag\\
&+& (0.284424n^2-0.684984n+1.642748)\varepsilon^2 \notag\\
&+& \left.\left. (0.03125n^2-0.0625+0.03125)\varepsilon^3\tilde{\varepsilon}^{-1}\right)\right].
\label{eta0}
\end{eqnarray}
In the case $n=1$, we have
\begin{equation}
\eta_0=-4[(9\tilde{\varepsilon}
 -6\varepsilon )/106]^{1/2}.  \label{eta0-n-1}
\end{equation}
Note that  Eq.~(\ref{eta0}) involves the ratio
$\varepsilon/\tilde{\varepsilon}$, which is of order
unity within our approximation. The authors of
Ref.~\onlinecite{prudnikov-83} pointed out that this ratio, which appears
in  $g^*$ and  $\gamma_{\tau}$, is absent from the final critical exponents
$\nu_{\perp}$, $\eta_{\perp}$, $z_{\perp}$, and $\zeta$ at second order.
This encouraged them to believe that similar cancellations will occur at
all orders. It is easy to see that the reason for the ratio cancellation
is that the last three exponents do not contain  the first-order term proportional
to $g^{*}$, which gives rise to the ratio $\varepsilon/\tilde{\varepsilon}$
at second order. In the case of $\nu_{\perp}$, the accidental cancellation of terms
$\varepsilon/\tilde{\varepsilon}$ occurs when one substitutes FP (\ref{fp})
into $\gamma_{\tau}$. However, the cancellation can be destroyed, for example,
by introducing cubic anisotropy, \cite{yamazaki-86} so that all exponents involve the
ratio $\varepsilon/\tilde{\varepsilon}$. For physical applications, we are interested
in positive values $\varepsilon$ and $\tilde{\varepsilon}=\varepsilon+\varepsilon_d$,
for which $0<\varepsilon/\tilde{\varepsilon}<1$, and therefore the critical exponent
(\ref{eta0}) is well defined.

In order to describe the scaling behavior of the nonequilibrium critical
relaxation, we employ a short-distance expansion of the fields $\phi$
and $\tilde{\phi}$ in terms of the initial fields\cite{janssen-89}
for $t\to 0$,
\begin{eqnarray}
 \phi^{\beta}(t)=\sigma(t)\dot{\phi}_0^{\beta}+\ldots, \ \ \
 \tilde{\phi}^{\beta}(t)=\tilde{\sigma}(t)\tilde{\phi}_0^{\beta}+\ldots.
 \label{expansions}
\end{eqnarray}
Introducing the renormalized amplitude functions according to
$\sigma_b(t)=Z_0^{-1/2}\sigma(t)$ and
$\tilde{\sigma}_b(t)=Z_0^{-1/2}\tilde{\sigma}(t)$, we can derive the corresponding
RG equations $[\hat{R}-\eta_0/2]\sigma(t)=0$ and
$[\hat{R}-\eta_0/2]\tilde{\sigma}(t)=0$. Taking into account
$[\sigma]\sim(\lambda\mu^2)^{-1}$ and $[\tilde{\sigma}]\sim 1$, we find the
solutions of the RG equations at FPs
\begin{eqnarray}
&&\mbox{}\hspace{-7mm}  \sigma (t,\tau,\lambda,\alpha;\mu)= l^{-z_{\perp}-\eta_0/2}
   \sigma (t l^{z_{\perp}},\tau l^{-1/\nu_{\perp}},\lambda,\alpha;\mu), \notag \\
&&\mbox{}\hspace{-7mm}  \tilde{\sigma} (t,\tau,\lambda,\alpha;\mu)= l^{-\eta_0/2}
   \tilde{\sigma} (t l^{z_{\perp}},\tau l^{-1/\nu_{\perp}},\lambda,\alpha;\mu).
\label{sigma-rg}
\end{eqnarray}
We are interested  in the short-time critical behavior of the response and correlation
functions. Inserting expansions (\ref{expansions}) into  appropriate Green functions,
we obtain
$G(\mathbf{x},t,t^{\prime})=\tilde{\sigma}(t^{\prime})G_{0,1}^{1}(\mathbf{x},t)+\ldots$
and
$C(\mathbf{x},t,t^{\prime})=2\lambda\sigma(t^{\prime})G_{0,1}^{1}(\mathbf{x},t)+\ldots$,
where we have used Eq.~(\ref{relation}).
Combining Eqs.~(\ref{scaling}) and (\ref{sigma-rg}), we obtain the scaling behavior of
the response and correlation functions for
$\xi_{\perp}^{-z_{\perp}}t^{\prime}\to 0$ and $\xi_{\perp}^{-z_{\perp}}t>0$,
\begin{eqnarray}
&&\mbox{}\hspace{-4mm} G(\mathbf{x},t,t^{\prime})={x}_{\perp}^{2-d-\eta_{\perp}-z_{\perp}}
   \left( \frac{t}{t^{\prime}}\right)^{\theta}F_R\left(
   \frac{{x}_{\perp}}{ \xi_{\perp}},
   \frac{{x}_{\perp}}{{x}_{\parallel}^{\zeta}},
   \frac{t}{\xi_{\perp}^{z_{\perp}}}   \right), \ \ \ \  \notag \\
&& \mbox{}\hspace{-4mm} C(\mathbf{x},t,t^{\prime})={x}_{\perp}^{2-d-\eta_{\perp}}
   \left( \frac{t}{t^{\prime}}\right)^{\theta-1}F_C\left(
   \frac{{x}_{\perp}}{ \xi_{\perp}},
   \frac{{x}_{\perp}}{{x}_{\parallel}^{\zeta}},
   \frac{t}{\xi_{\perp}^{z_{\perp}}}   \right),\ \ \ \ \label{GC}
\end{eqnarray}
where the new dynamic critical exponent is defined by
\begin{equation}
\theta=-\eta_0/(2z_{\perp}). \label{scal-rel0}
\end{equation}
To second order in $\varepsilon$ and $\tilde{\varepsilon}$, it reads
\begin{eqnarray}
\theta &=& (n+2)(n-1)^{-1} \left[ 0.1875\tilde{\varepsilon}- 0.125 \varepsilon
  -  (n-1)^{-2} \right. \notag \\
&\times & \left( (0.062119 n^2 + 0.232311n + 0.404301) {\tilde{\varepsilon}}^{2}\notag \right.\\
&-&(0.169556 n^2 - 0.021917n + 0.784002) \varepsilon \tilde{\varepsilon} \notag\\
&+&(0.007813 n^2 + 0.007813n - 0.015625) {\varepsilon}^{3} \tilde{\varepsilon}^{-1} \notag\\
&+& \left.\left. (0.086731 n^2 - 0.155621n + 0.379437) {\varepsilon}^{2}
\right)\right].
\label{thet}
\end{eqnarray}
Although the system is characterized by the single exponent $\theta$, the scaling laws
(\ref{GC}) reflect the strong anisotropy of the system under consideration.

%%%%%%%%%%%%%%%%%%%%%%%%%%%%%%%%%%%%%%%%%%%%%%%%%%%%%%%%%%%%%%%%%%%%%%%%%%%%%%%%%%%%
\begin{table}[tbp]
\caption{Values of critical exponents $\eta_0$, $\theta$, and $\theta^{\prime}$ for $d=3$.}
\label{table2}%
\begin{ruledtabular}
\begin{tabular}{lllll}
 $n$  & $\varepsilon_d$   &  $\eta_0$   &  $\theta$     &  $\theta^{\prime}$\\ \hline
  1   & 0\footnotemark[1] &  $-0.374$   &   $0.092$     &  $0.087$          \\
      &       1           &  $-1.346$   &   $0.252$     &  $0.140$          \\
      &       2           &  $-1.780$   &   $0.308$     &  $0.331$          \\
  2   & 0\footnotemark[2] &  $-0.647$   & $0.162$ ($0.160$)    &  $0.200$          \\
      &       1           &  $-0.614$   & $0.142$ ($0.138$)    &  $0.059$          \\
      &       2           &  $-0.688$   & $0.157$ ($0.152$)    &  $0.054$          \\
  3   & 0\footnotemark[2] &  $-0.790$   & $0.198$ ($0.196$)    &  $0.229$          \\
      &       1           &  $-0.834$   & $0.191$ ($0.186$)    &  $0.111$          \\
      &       2           &  $-1.008$   & $0.226$ ($0.214$)    &  $0.111$          \\
\end{tabular}
\end{ruledtabular}
\footnotetext[1]{Taken from Ref.~\onlinecite{oerding-95}. }
\footnotetext[2]{Computed using the results of Ref.~\onlinecite{janssen-89} for pure systems.}
\end{table}
%%%%%%%%%%%%%%%%%%%%%%%%%%%%%%%%%%%%%%%%%%%%%%%%%%%%%%%%%%%%%%%%%%%%%%%%%%%%%%%%%%%%

We now discuss the scaling behavior of the order parameter $m^{\beta}(\mathbf{x},t)$
as a function of its initial value $m^{\beta}_0(\mathbf{x})$. For the sake of clarity,
we will suppress the superscript $\beta$ in what follows. Considering $m_0(\mathbf{x})$ as an
additional time-independent source coupled to the field $\tilde{\phi}_0(\mathbf{x})$,
we add the term $\int d^dx\, m_0(\mathbf{x}) \tilde{\phi}_0(\mathbf{x})$ to the effective
action (\ref{eq:L}). The order parameter is given then by
\begin{equation}
m(\mathbf{x},t)=\overline{\langle\phi(\mathbf{x},t)\rangle}=
\delta W[h,\tilde{h}]/\delta h(\mathbf{x},t)|_{h=\tilde{h}=0}.
\label{order-param}
\end{equation}
Expanding Eq.~(\ref{order-param}) in $m_0$ we obtain for the
case of a homogeneous initial order parameter
\begin{equation}
m(m_0,t)=\sum\limits_{L=1}^{\infty}\frac1{L!}\int d^dx_1\ldots\int d^dx_{L}
G^{L}_{0,1}(\{\mathbf{x}\},t)m_0^{L}.
\label{m-m0}
\end{equation}
Equation~(\ref{m-m0}) holds also for the renormalized quantities, so that
substituting Eq.~(\ref{scaling}) into Eq.~(\ref{m-m0}) and summing
over $L$, we obtain
\begin{equation}
m(m_0,t)= m_0t^{\theta^{\prime}}F(m_0 t^{\theta^{\prime}
+\beta/\nu_{\perp}z_{\perp}},\tau t^{1/\nu_{\perp}z_{\perp}}),\label{order}
\end{equation}
where $\theta^{\prime}=-[\eta_{\perp}+\tilde{\eta}+\eta_{0}+%
2\varepsilon_d(1-\zeta)]/(2z_{\perp})$, $\beta$ is the critical exponent
for the order parameter and we have used the scaling relation
$2\beta=\nu_{\perp}(\eta_{\perp}-2+\tilde{d})+\varepsilon_d\nu_{\parallel}$.
\cite{lawrie-84} For $n>1$, we have
\begin{eqnarray}
\theta^{\prime} &=& 0.125 \tilde{\varepsilon} + (n-1)^{-2} \notag \\
& \times &\left[(0.058228 n^2 - 0.062655n + 0.204281) \varepsilon
   \tilde{\varepsilon}\right. \notag\\
&+& (0.024597 n^2 + 0.055126n + 0.011862) \varepsilon^2 \notag\\
&-& (0.007813n^2+ 0.03125n + 0.03125) \varepsilon^3 \tilde{\varepsilon}^{-1}
     \notag\\
&-& \left. (0.034287 n^2 + 0.088644n + 0.145612) \tilde{\varepsilon}^2 \right].
\label{thet1}
\end{eqnarray}
The critical exponents $\theta$ and $\theta^{\prime}$ are
related by
\begin{equation}
\theta^{\prime}=\theta+(2-\eta_{\perp}-z_{\perp}-\varepsilon_d(1-\zeta))/z_{\perp}.
\label{scal-rel}
\end{equation}
The scaling function $F(y_1,y_2)$ has
a universal behavior at the critical point $\tau=0$:
$F(0,0)$ is finite, while for $y_1\to \infty$, $F(y_1,0)\propto 1/y_1$.
Thus, Eq.~(\ref{order}) exhibits the crossover between
the initial and the equilibrium stages of the critical relaxation at time
$t_c\propto m_0^{-1/(\theta^{\prime}+\beta/\nu_{\perp}z_{\perp})}$, when the
initial increasing  $m(t)\propto t^{\theta^{\prime}}$ turns
to the long-time decreasing $m(t)\propto t^{-\beta/\nu_{\perp}z_{\perp}}$.

It is well known that the convergence of double series in $\varepsilon$
and $\tilde{\varepsilon}$ is poor.\cite{lawrie-84,blavatska-03}
For $n=1$, we know only the lowest order term (\ref{eta0-n-1})
and cannot apply any resummation technique. Therefore, we compute
exponent (\ref{eta0-n-1}) as it is, and then we calculate $\theta$ and
$\theta^{\prime}$ by using Eqs.~(\ref{scal-rel0}) and (\ref{scal-rel}).
In order to estimate the values of the critical exponents $\eta_0$, $\theta$,
and $\theta^{\prime}$ for $n>1$, we employ the Pad\'{e}-Borel resummation
method extended to the two-parameter case (see Ref.~\onlinecite{holovatch-02}
and references therein).
Numerical values of the critical exponents $\eta_0$, $\theta$, and $\theta^{\prime}$
computed for the three-dimensional Ising ($n=1$), XY ($n=2$), and Heisenberg ($n=3$)
systems with pointlike ($\varepsilon_d=0$), linear ($\varepsilon_d=1$), and
planar ($\varepsilon_d=2$) defects are presented in Table~\ref{table2}.
To calculate $\theta$ and $\theta^{\prime}$, we can use either expansions (\ref{thet})
and (\ref{thet1}), or scaling relations  (\ref{scal-rel0}) and (\ref{scal-rel}).
This can be utilized to check the accuracy of the resummation procedure.
The values of $\theta$ given in parenthesis (see Table~\ref{table2}) are computed
with the use of Eq.~(\ref{scal-rel0}). The small discrepancies indicate the fairly
good accuracy of the computed values of $\theta$.
Unfortunately, using the scaling relation (\ref{scal-rel}) leads to
unreasonable values of $\theta^{\prime}$. The reason is the large error in the
determination of $\eta_{\perp}$. Indeed, the direct calculation of $\eta_{\perp}$
gives $\eta_{\perp}\approx-0.44$ for $n=2$ and $\varepsilon_d=1$,\cite{lawrie-84}
while the scaling relation $\eta_{\perp}=2-\gamma/\nu_{\perp}$, where $\gamma$ is the
critical exponent of susceptibility, suggests $\eta_{\perp}\approx0.008$.
\cite{blavatska-03} The resummation of expansion (\ref{thet1}) results in more
plausible values of $\theta^{\prime}$, which are shown in Table~\ref{table2}.

According to the Harris criterion,\cite{harris-74} the pointlike disorder ($\varepsilon_d=0$)
is relevant for the critical behavior of  three-dimensional systems
only if $n=1$. The extended defects change the critical behavior for all $n$
if $\varepsilon_d>\varepsilon_d^{\textrm{marg}}(n)$, where the marginal values
$\varepsilon_d^{\textrm{marg}}(1)=-0.173$, $\varepsilon_d^{\textrm{marg}}(2)=0.016$,
and $\varepsilon_d^{\textrm{marg}}(3)=0.172$  were calculated for $d=3$ in
Ref.~\onlinecite{blavatska-03} using the extended Harris criterion (\ref{harris}).
As one can see from Table~\ref{table2},  the 3D Ising system
with extended defects is characterized by larger values of the initial slip
exponents $\theta$ and $\theta^{\prime}$ than the 3D Ising system with
pointlike disorder.
For $n>1$, the pointlike defects do not affect the critical behavior and
the critical exponents take  the same values as that of the pure system.
Table~\ref{table2} shows that for $n=2,3$,
the presence of extended defects leads to the small changes in $\theta$, while
the corresponding decrease of $\theta^{\prime}$ is more significant and can be
observed in numerical simulations. Remarkably, the decrease of $\theta^{\prime}$
caused by extended defects almost does  not depend on $\varepsilon_d$ for $n>1$
in the considered approximation.

%%%%%%%%%%%%%%%%%%%%%%%%%%%
\section{Conclusions}
%%%%%%%%%%%%%%%%%%%%%%%%%%%

In the present paper, we have studied the short-time critical  dynamics of the
system with extended defects, which are strongly correlated in $\varepsilon_d$
dimensions and randomly distributed over the remaining $\tilde{d}%
=d-\varepsilon_d$ dimensions. We have considered the nonequilibrium relaxation
of the system starting from a macroscopically prepared initial state with short-range
correlations and found the scaling behavior of the nonequilibrium correlation and
response functions, and the initial increase of the order parameter.
Using the Pad\'{e}-Borel resummation method, we have calculated the initial slip
exponents $\theta$ and $\theta^{\prime}$, which describe the initial growth of
correlations, to two-loop order.

Unfortunately, to our knowledge there is only one numerical investigation of
the short-time critical dynamics of systems with extended defects.
In Ref.~\onlinecite{zheng-02}, the short-time
critical dynamics of the 2D Ising model with bond disorder perfectly
correlated in one dimension and uncorrelated in the
other was studied using Monte Carlo simulation.
This model corresponds to the case $\varepsilon=2$ and $\varepsilon_d=1$,
for which Eq.~(\ref{scal-rel}) gives $\theta^{\prime}=0.154$.
The value $\theta^{\prime}=0.176(2)$ reported in
Ref.~\onlinecite{zheng-02} is larger than the prediction
of the RG analysis. The difference can be attributed to
very strong disorder used in Ref.~\onlinecite{zheng-02}
(the concentration of bonds $p=0.5$), while the RG description
is valid only for the regime of weak disorder ($p$ closes to 1), in
which the asymptotic behavior does not depend on the concentration
$p$.\cite{parisi-99}

The relevant question is  how  the critical properties of the  system are
modified if the defects, oriented along the direction  $\mathbf{x}_{\parallel}$,
are very long but finite, instead of being extended throughout the system.
According to the usual scaling  theory near the critical point $T_c$, the only
relevant scale that remains in the system is the correlation length, which
diverges in the critical point as $\xi\propto|\tau|^{-\nu_p}$.  If the disorder is weak,
its effect on the critical behavior in the vicinity of the critical point is negligible
as long as the correlation length $\xi$ is smaller than the average distance between
the defects $l_{\rm def}$, i.e., the critical behavior is controlled by the FP of the pure system.
In close vicinity of the critical point ($\tau\to 0$), the correlation length
grows and becomes large than  $l_{\rm def}$ for $\tau\ll \Delta^{-1/\varphi}$,
where $\varphi$ is the crossover exponent (\ref{harris}).
However, it is possible that the correlation length remains smaller than the length of defects.
For these values of $\tau$, the critical behavior is controlled by the FP of a
system  with extended defects, the two correlation
lengths naturally arise due to the strong anisotropy, and one can use the results obtained here
for the case of the strong correlated disorder. Closer to the critical point, the
correlation length $\xi_{\parallel}$ becomes larger than the length of defects and the asymptotic
behavior exhibits crossover to the critical behavior of the system
with pointlike disorder.
The corresponding crossover exponent is $\nu_{\parallel}$.

From an experimental point of view,  the considered model may be relevant for the anomalous
scaling behavior of the so-called narrow component in magnetic systems such as Ho and Tb,
which exhibit two different length scales for critical fluctuations usually ascribed
to long-range correlated quenched disorder.\cite{altarelli-95} Other possible
applications are high-$T_c$ superconducting  materials, where extended disorder is caused
by irregularly alternating layers, such as those containing Y and Ba in Y-Ba-Cu-O structures,
or CuO chains. \cite{decesare-94}
The obtained results can be used to investigate the
long-time dynamics of systems with extended defects in the aging regime.\cite{calabrese-02}
They  can be helpful also for numerical simulations of systems
with extended defects, especially by using  new effective
algorithms, based on the fact that due to the small correlation length
during the initial stage of the critical relaxation, the determination of
critical exponents requires less effort than numerical studies of
equilibrium dynamics. \cite{li-95}

\begin{acknowledgments}
The support from the Deutsche Forschungsgemeinschaft (SFB 418) is gratefully
acknowledged. I would also like to thank Y. Chen for a useful discussion.
\end{acknowledgments}

\end{document}